\documentstyle[aps,epsf,floats]{revtex}
\begin{document}

\begin{flushright}
JINR E2-82-159 \\
February 26,1982\\
JINR Rapid Communications \\
4[78]-96,9 (1996) 
\end{flushright}

\vspace{1cm}

\begin{center}
{\Large \bf OPTIMIZED LAMBDA-PARAMETRIZATION  \\
FOR THE QCD RUNNING COUPLING CONSTANT \\ IN SPACELIKE AND TIMELIKE 
REGIONS$^*$}

\end{center}

\vspace{1cm}

\centerline{\Large \bf A.V. Radyushkin$^{**}$}

\begin{center}
{\em $^*$The  investigation has been performed 
(and completed in February 1982) 
at the  } \\ {\em
  Laboratory of Theoretical  Physics, 
JINR, Dubna, Russian Federation}

\vspace{1cm}

{\em $^{**}$Present address: 
Physics Department, Old Dominion University,
Norfolk, VA 23529, USA} \\ {\em and} \\{\em  Theory Group, Jefferson Lab,
Newport News, VA 23606, USA}

\end{center} 

\begin{abstract}

The algorithm is described that enables one
to perform an explicit summation of all the 
($\pi^2/\ln (Q^2/\Lambda^2))^N$ 
corrections to $\alpha_s (Q^2)$ that appear owing 
to the analytic continuation
from spacelike to timelike region of momentum transfer.

\end{abstract}

\newpage

\centerline{\bf I. INTRODUCTION}

\vspace{5mm}

Perturbative QCD  is intensively applied now \{1\} to various processes 
involving large momentum transfers, both in spacelike
 $(q^2 = -Q^2 < 0)$ and timelike $(q^2 >0)$ 
regions (for a review see \cite{1}$-$\cite{3}). However,
the coupling constant $g(\mu)$ (i.e., the expansion parameter)
is defined usually with the reference to some
Euclidean (spacelike) configuration of momenta of scale $\mu$.
For spacelike $q$ this produces no special complications.
One simply uses the renormalization group to sum up
the logarithmic corrections  $(g^2(\mu) \ln (Q^2/ \mu^2))^N$
that appear in higher orders and arrives at the expansion
in the effective coupling constant $\alpha_s (Q^2)$ which in the lowest
approximation is given by the famous asymptotic freedom
formula \cite{1}
\begin{equation}
\alpha_s (Q^2) = \frac{4 \pi}{(11-2N_f/3) \ln (Q^2/ \Lambda^2)} \ ,
\label{1}
\end{equation}
where $\Lambda$ is the ``fundamental" scale of QCD. In general,
the $\Lambda$-parametrization of $\alpha_s (Q^2)$
is a series expansion in $1/L$ (where $L= \ln (Q^2/ \Lambda^2)$ ),
and the definition of $\Lambda$ is fixed only if 
the $O(1/L^2)$-term is added to Eq.(1) \cite{4}.

For timelike $q$ there appear, however, $i \pi$-factors ($\ln (Q^2/ \mu^2)
\to \ln (Q^2/ \mu^2) \pm i \pi$), and it is not clear a priori
what is the effective expansion parameter in this region.
This problem was discussed recently \{1\} in a very suggestive paper by
Pennington and Ross \cite{5}. These authors analyzed the ratio
$R(q^2) = \sigma (e^+ e^- \to {\rm hadrons})/ \sigma (e^+ e^- \to \mu^+ \mu^-)$
for which the analytic continuation  from the spacelike
to timelike region is well-defined and investigated which
of the three ans{\"a}tze 
($\alpha_s (q^2), |\alpha_s (-q^2)| $ and ${\rm Re}\, \alpha_s (-q^2) $  )
better absorbs the $(\pi^2/L^2)^N$-corrections\footnote{Odd powers
of $(i\pi/L)$ cancel because $R$ is real.} 
in the timelike region $q^2 >0$.
Their conclusion was that $|\alpha_s (-q^2)|$ is better than $\alpha_s (q^2)$ 
and 
${\rm Re}\, \alpha_s (-q^2) $. Nevertheless, it is easy
to demonstrate by a straightforward calculation
that $|\alpha_s (-q^2)|$  cannot absorb all the $(\pi^2/L^2)^N$-terms
associated with the analytic continuation of the $\ln (Q^2/ \mu^2)$-factors.
Our main goal in the present letter is to show that by using 
the $\Lambda$-parametrization for $\alpha_s (Q^2) $ in the spacelike
region it is possible to construct for
$R(q^2)$ in the timelike region the expansion in which 
all  the $(\pi^2/L^2)^N$-terms are summed explicitly.

\vspace{5mm}

\centerline{\bf II. $\Lambda$-PARAMETRIZATION IN SPACELIKE REGION}

\vspace{5mm}

The starting point for the $\Lambda$-parametrization
is the Gell-Mann-Low equation taken as a series expansion 
in $G= \alpha_s/4 \pi$:
\begin{equation}
L \equiv \ln (Q^2/\Lambda^2) = \frac1{b_0 G} +
\frac{b_1}{b_0^2 } \ln G + \Delta + \frac{b_2b_0 - b_1^2}{b_0^3}\, G
+O(G^2) \  , 
\end{equation}
where $b_k$ are $\beta$-function coefficients:\\

 $b_0=11-2N_f/3$ \cite{1},
$b_1=102-38N_f/3$ \cite{6}, $b_2^{MS} =2857/2-5033N_f/18 + 325 N_f^2/54$
\cite{7}.\\

 The parameter $\Delta$ in Eq.(2) is due to the lower boundary of
the GML integral \cite{8,9}.
By a particular choice of $\Delta$  one fixes the definition
of $\Lambda$:   $\Lambda= \Lambda(\Delta)$ \footnote{Of course,
$\Lambda$ depends also on the renormalization scheme 
chosen.}.  Eq.(2)  is solved by iterations and the 
result is reexpanded in $1/L$:
\begin{equation}
\alpha_s (Q^2) = \frac{4 \pi}{b_0 L} \left \{
1- \frac{L_1}{L} + \frac1{L^2} \left [ L_1^2 - \frac{b_1}{b_0^2 } L_1
+ \frac{b_2b_0 - b_1^2}{b_0^4} \right ] + O(1/L^3) \right \}\, ,
\end{equation}
where 
\begin{equation}
L_1 = \frac{b_1}{b_0^2 } \ln (b_0 L) - \Delta \ .
\end{equation}

The expansion (3) is useful, of course, only if it converges
rapidly enough. In fact, the convergence of the $1/L$ series
depends $ (i)$  on the value of $L$ we are interested in 
and $(ii)$ on the choice of $\Delta$.

We emphasize that the most important for perturbative QCD 
is the region $L>3$, since $L=3$ corresponds to
$\alpha_s \sim 0.5$, and the reliability of perturbation theory
for larger $\alpha_s$ is questionable. Hence, in a realistic situation 
the naive expansion parameter $1/L$ is smaller than (but usually close to)
one third. Of course, 1/3 is not very small, so one must check the 
coefficients of the $1/L$ expansion more carefully.
First, there is a $\Delta$-convention-independent 
term $(b_2 b_0 -b_1^2)/(b_0^4L^2)$ which reduces for $N_f = 3$ to roughly
$0.25/L^2$ and gives, therefore, less than 3$\%$-correction to the simplest 
formula (1). There are also $\Delta$-dependent terms like $L_1/L,
L_1/L^2$ and one should choose $\Delta$ so as to minimize the upper value 
of the ratio $L_1/L$ in the $L$-region of  interest.

If one takes, e.g., $$\Delta = \Delta_{opt} = (b_1/b_0^2) \ln (4 b_0)$$
then $L_1 = (b_1/b_0^2) \ln (L/4)$ and the ratio $L_1/L$ is smaller
than $7 \%$ in the whole region $L>3$. Another choice \cite{10}
is to take $$\Delta = \Delta (Q_0^2) = (b_1/b_0^2) \ln (b_0 L_0) ß , $$
where $L_0 = \ln (Q_0^2/\Lambda^2)$ and $Q_0^2$ lies 
somewhere in the middle
of the $Q^2$-region analyzed. In this case 
\mbox{$L_1 = (b_1/b_0^2) \ln
(L/L_0)$,} i.e.,  $L_1/L$ is zero for $Q^2 = Q_0^2$ and smaller than $7\%$ 
for all $Q^2$ in the region where $L>3$. 
An important observation is that both the choices minimize
the corrections not only in Eq. (3) but also in the GML equation (2).

Really, for small $G$ the only dangerous term in Eq. (2) is $\ln G$,
hence, the best thing to do is to compensate it by taking
$\Delta = - (b_1/b_0^2) \ln \overline G $, where $ \overline G$ is $\alpha_s
(Q^2)/4\pi$ averaged (in some sense) over the relevant $Q^2$-region.
After this has been done, one may safely solve Eq. (2) by iterations
and perform the $1/L$-expansion. For a proper choice
of $\Delta$ Eq. (3) has $1 \%$ accuracy for $L>3$, and, moreover,
the total correction to the simplest formula (1) is less than $10 \%$.
However, accepting the most popular prescription 
$$\Delta_{pop} = (b_1/b_0^2) \ln (b_0) = \Delta (Q_0^2 = e \Lambda^2)$$
(the only motivation for $\Delta_{pop} $ being the ``aesthetic"
criterion that $L_1$  should have the shortest form 
\mbox {$L_1 = (b_1/b_0^2) \ln (L)$ ) } one minimizes $L_1/L$ in the region
$Q^2 \sim 3 \Lambda^2$ nobody is really interested in.
Moreover, in the important region $L \sim 3$ one has $L_1^{pop}/L \sim
1/3$, and the convergence of the $1/L$ series is very poor
in this case.

Thus, the $\Lambda$-parametrization (Eq.(3)) gives a
 rather compact and
sufficiently precise expression for the effective 
coupling constant in the spacelike region provided a proper
choice of the $\Delta$-parameter has been made.

\vspace{5mm}

 \centerline{\bf III. $\Lambda$-PARAMETRIZATION  AND 
${\bf R(e^+e^- \to {\rm \bf
hadrons}, s)}$}

\vspace{5mm}

The standard procedure (see, e.g., \cite{11} and references therein)
is to calculate the derivative $D(Q^2) = Q^2 dt/dQ^2$ of the vacuum
polarization $t(Q^2)$ related to $R$ by
\begin{equation}
R(s) = \frac1{2 \pi i} \left ( t(-s+i\epsilon) - t(-s-i \epsilon) \right ) \ .
\end{equation}
In perturbative QCD $D(Q^2)$ is given by the $\alpha_s (Q^2)$-expansion:
\begin{equation}
D(Q^2) = \sum_q e_q^2 \left \{ 1 + \frac{\alpha_s(Q^2)}{\pi}
+ d_2 \left (\frac{\alpha_s(Q^2)}{\pi} \right )^2 
+d_3 \left (\frac{\alpha_s(Q^2)}{\pi} \right )^3 + \ldots \right \} \ .
\end{equation}

Only $d_2$ is known now \cite{11,12} \{2\}, its value depending on 
the renormalization scheme chosen. Using Eq. (5) and the definition 
of $D$, one can relate $R(s)$ (or, more precisely, its
perturbative QCD version $R^{QCD} (s)$ ) directly
to $D(Q^2)$
\begin{equation}
R^{QCD} (s) = \frac1{2\pi i} \int_{-s-i\epsilon}^{-s+i\epsilon}
D(\sigma) \frac{d\sigma}{\sigma} \ .
\end{equation}
Integration in Eq.(7) goes below the real axis from
$-s-i\epsilon$
to zero \{3\} and then above the real axis to $-s+i\epsilon$. 

In a shorthand notation $ D \Rightarrow  R \equiv \Phi [D]$.
In some important cases the integral can be calculated explicitly \{4\} :
\begin{eqnarray}
 & & 1 \Rightarrow 1 \\
& &\frac1{L_{\sigma}} \Rightarrow 
\frac1{\pi} \arctan (\pi/L_s) = \frac1{L_s} \left \{1- 
\frac13 \frac{\pi^2}{L_s^2} + \ldots \right \} \\& &
\frac{\ln (L_{\sigma}/L_0)}{L_{\sigma}^2} \Rightarrow 
\frac{\ln (\sqrt{L_s^2+\pi^2}/L_0) - (L_s/\pi)\arctan (\pi/L_s) +1 }
{L_s^2+\pi^2}  = 
\frac{L_s/L_0}{L_s^2} \left \{1- 
\frac{\pi^2}{L_s^2} + \ldots \right \} + \frac56 \frac{\pi^2}{L_s^4} + \ldots 
\ \,  \\& &
\frac1{L_{\sigma}^2} \Rightarrow \frac1{L_s^2+\pi^2} = \frac1{L_s^2} \left \{1- 
\frac{\pi^2}{L_s^2} + \ldots \right \} \ \,  \\& &
\frac1{L_{\sigma}^n} \Rightarrow (-1)^n \frac1{(n-1)!}
\left ( \frac{d}{dL_s} \right )^{n-2} \frac1{L_s^2+\pi^2} =
\frac1{L_s^n} \left \{1-  \frac{\pi^2}{L_s^2} \frac{n(n+1)}{6}+ \ldots \right
\} \ \,  
\end{eqnarray}
where $L_s= \ln (s/\Lambda^2), L_{\sigma} = \ln (\sigma /\Lambda^2)$
and $L_0$ is a  constant depending  on the $\Delta$-choice.

Using the $\Lambda$-parametrization for $\alpha_s (\sigma)$ 
and incorporating Eqs.(8)-(12) (as well as their
generalizations for $\ln^2 L/L^3, \ln L/L^3$ etc.) produces
the expansion for $R^{QCD} (s)$ 
\begin{equation}
R^{QCD} (s) = \sum_q e_q^2 
\left \{ 1+ \sum_{k=1} d_k \Phi [(\alpha_s/\pi)^k] \right \}
\end{equation}
in which all the $(\pi^2/L^2)^N$-terms are summed up explicitly.

\vspace{5mm}

\centerline{\bf IV.  QUEST FOR THE BEST EXPANSION PARAMETER}

\vspace{5mm}

Note that the expansion (13) is not an expansion 
in powers of some particular parameter since the application of the
$\Phi$-operation normally violates nonlinear relations:
$\Phi [1/L^2] \neq  (\Phi [1/L])^2$, etc. 
A priori, there are no grounds to believe that a power expansion is better
than any other (say, Fourier).
In fact, the expansion (13) converges better than the
generating expansion (6) for $D(\sigma)$ because,
as it follows from   Eqs. (9)-(12), 
$\Phi [\alpha_s^N]$ is always smaller than $\alpha_s^N$.
Moreover, $(\Phi [\alpha_s^{N+1}])^{1/(N+1)} <
(\Phi [\alpha_s^{N}])^{1/N} $, i.e., the effective expansion parameter
decreases in higher orders.
Thus, if one succeeded in obtaining a good $\alpha_s^{N}$
expansion for $D(\sigma)$ (with all $d_N$ being small numbers),
then the resulting $\Phi [\alpha_s^{N}]$ expansion 
for $R^{QCD}(s)$ is even better, and the best thing to do is to 
leave it as it is.

However, if one insists that the result for $R^{QCD}(s)$ 
should have a form of a power expansion, then the best
expansion parameter is evidently $\Phi [\alpha_s/\pi]$ 
because the largest nontrivial (i.e., $O(\alpha_s/\pi)$ \, )
term of the expansion is reproduced in the exact form and only higher
terms are spoiled. The analogue of the simplest 
$\Lambda$-parametrization for $\alpha_s (Q^2)$ (Eq.(1) ) is then
\begin{equation}
\tilde \alpha_s (q^2) = \frac{4}{b_0} \arctan
 \left ( \frac{\pi}{\ln (q^2/\Lambda^2)} \right ) \ .
\end{equation}

Using Eqs. (8)-(13) it is easy to realize that 
$\alpha_s (q^2) $ is really a bad expansion parameter,
because if one reexpands $\tilde \alpha_s (q^2) $ in 
$\alpha_s (q^2) $, there appear terms with large coefficients
\begin{equation}
\tilde \alpha_s (q^2) = \alpha_s (q^2) \left \{
1- \frac13 \left ( \frac{\pi b_0}{4} \right )^2
 \left ( \frac{\alpha_s (q^2)}{\pi} \right )^2  + \ldots \right \}
\approx \alpha_s  \left \{1- 17  \left ( \frac{\alpha_s }{\pi} \right )^2 
 + \ldots \right \} \ .
\end{equation}

If one reexpands $\tilde \alpha_s (q^2) $ in Re\, $\alpha_s (-q^2)$
then the corresponding coefficient is even 2 times larger,
whereas if  $\tilde \alpha_s (q^2) $ is reexpanded in 
$|\alpha_s (-q^2)|$, the coefficient is 2 times smaller.
This observation is in full agreement with the result
of ref. \cite{5} quoted in the Introduction.

\vspace{5mm}

\centerline{\bf V. CONCLUDING REMARKS}

\vspace{5mm}

It should be noted that the change of the expansion 
parameter as given by Eq. (15) affects only the $(\alpha_s /\pi)^3$
coefficient of the $R^{QCD}$-expansion which has not  been calculated 
yet
\{2\}. So, within the present-day accuracy, all  expansions
for $R^{QCD}$ have the same  coefficients.
It is worth emphasizing, nevertheless,
that the $\pi^2/L^2$ terms produce for 
$\alpha_s \gtrsim 0.3$ more than 20\% correction
to $\alpha_s$, i.e., they are more important (for an optimal
choice of the $\Delta$-parameter) than the 2-loop corrections in Eq. (3).

To conclude, we have described the construction of an optimized (i.e.
rapidly convergent) $\Lambda$-parametrization 
for the effective QCD coupling   constant 
in the spacelike region, and then we used it to obtain the fastest convergent
expansion for the timelike quantity  $R^{QCD}(s)$.
The technique outlined in the present paper
can be applied also to other  $R^{QCD}$-like quantities.
Such quantities do appear, e.g., in the QCD sum rule
approach \cite{13} in which the analysis of hadronic properties
is based on the study of vacuum correlators
of   various currents. They appear also in an alternative approach
\cite{14} based on the finite-energy sum rules \cite{15}.
It should be stressed that in the latter approach 
the   $R^{QCD}$-like quantities enter into the basic integral relation,
and the analysis is most conveniently performed if 
one has a simple analytic expression similar to that described above. 

\vspace{5mm}

\centerline{\bf ACKNOWLEDGEMENTS}

\vspace{5mm}

I am grateful to A.V. Efremov and 
D.V. Shirkov for their interest in this work and support.
I thank D.I. Kazakov,
O.V. Tarasov  and A.A. Vladimirov for useful 
discussions.

\centerline{\bf  NOTES ADDED}

\vspace{5mm}

 \{1\} This paper was submitted in 1982 to Physics Letters B, but not 
accepted because the referee was not convinced that it needs a rapid 
publication. I was  recommended to write a longer version and submit it 
to a regular journal. Unwisely, I did not do that.
Still, though the paper  existed  in the preprint form only, it 
was known to experts in multiloop calculations. In particular, 
Eq.(15) was incorporated  into  the 4-loop   calculation of
$R^{QCD}(s)$  [16,17].
 Later, some of my results were used or 
rediscovered in several publications 
including very  recent ones (see, e.g.,[18-22]).
In 1996, the paper was reprinted in JINR Rapid Communications [23],
but the subsequent experience convinced me 
that the only way to make the paper accessible
to interested readers (and hopefully get credit
for its results) is to submit it to an e-print archive.
The original  version is reproduced above without any changes.
I only added references 
to short notes presented below. They contain updating 
and clarifying remarks.

\{2\}  The coefficient $d_3^{\overline{MS}}$ was calculated 
in refs. [16,17].  Starting from 
 this level, the coefficients $r_k$ of the $\alpha_s(s)$-expansion
for $R^{QCD}(s)$  
$$R^{QCD}(s)= \sum_q e_q^2 \left \{ 1 + \frac{\alpha_s(s)}{\pi}
+ r_2 \left (\frac{\alpha_s(s)}{\pi} \right )^2 
+r_3 \left (\frac{\alpha_s(s)}{\pi} \right )^3 + \ldots \right \} 
$$ 
differ from $d_k$:
according to our Eq.(15)    $r_3= d_3 - (\pi b_0)^2/48$.

 \{3\} This is an 
 incorrect description of the actual straightforward 
procedure which I used  to get results displayed in Eqs. (8)-(12). 
The central  idea of the paper is to represent $D(\sigma)$
as a sum of terms for each of  which the integral (7) can be calculated 
as an explicit analytic expression like 
$\ln \ln \sigma$ and then simply take the difference 
of these integrals at $-s+i\epsilon$ and $-s - i\epsilon$
using the standard prescription that the cut of $\ln z$
is on the negative real axis.
This is precisely what one should do to   
 analytically continue $t(Q^2)$ from the deep spacelike region. 
However, for terms in $ D(\sigma)$ containing  poles at $\sigma = \Lambda^2$
this prescription  is equivalent to integration
from $-s-i\epsilon$ below the real axis to some real 
point $\sigma_0 >
\Lambda^2$ (rather than zero) and then above
 the real axis to $-s+i\epsilon$. 

\{4\} In 1982, the applications of 
perturbative QCD  in the $0<s<\Lambda^2$
region were not treated as reliable,
 so   $s> \Lambda^2$ (i.e. $L_s>0$)
is implied  in Eqs. (8)-(12). Furthermore,  the results are  
presented in a form  most suitable  for the $1/L_s$ expansion.
To reconstruct  the original expressions valid  both for positive and
negative $L_s$ one should  change 
$$ \arctan (\pi/L_s) \to \pi/2 - \arctan (L_s/\pi) $$
in Eqs. (8)-(12), but this form is not very illuminative 
 for large $L_s$.

\vspace{5mm}

\centerline{\bf REFERENCES ADDED}

\vspace{5mm}

[16]  S.G. Gorishnii, A.L. Kataev, S.A. Larin,
Phys.Lett. B259, 144 (1991).

[17] L.R. Surguladze,  M. A. Samuel,
Phys.Rev.Lett. 66, 560 (1991).

[18] A.A. Pivovarov, Nuovo Cim. 105A, 813 (1992).

[19] H.F. Jones, I.L. Solovtsov, Phys.Lett. B349, 519 (1995), 
 hep-ph/9501344. 

[20] K.A. Milton, O.P. Solovtsova, Phys.Rev. D57, 5402 (1998),
 hep-ph/9710316.

[21] B.V. Geshkenbein, B.L. Ioffe, hep-ph/9906406.

[22] D.V. Shirkov, I.L. Solovtsov, hep-ph/9906495. 

[23] A.V. Radyushkin, JINR Rapid Comm. 4[78]-96,9 (1996).

\end{document}